\documentclass[twocoloumn]{aa} 

\usepackage{graphicx}
\usepackage{txfonts}
\usepackage{color}
\usepackage[colorlinks]{hyperref}
\hypersetup{ colorlinks, linkcolor=blue, citecolor=blue }

\newcommand{\Msun}{\ensuremath{M_{\sun}}}

\defcitealias{Han2004}{HP04}

\begin{document}

\title{A closer look at non-interacting He stars as a channel for producing the old population of type I\lowercase{a} supernovae}

\titlerunning{A particular MS donor channel to the old SNe Ia}
\authorrunning{Z.-W. Liu \& R. J. Stancliffe}

   \author{Zheng-Wei Liu\inst{1,2,3}
           \and 
          Richard J. Stancliffe\inst{4}
          }

   \institute{
Yunnan Observatories, Chinese Academy of Sciences (CAS), 396 Yangfangwang, Guandu District, Kunming 650216, P.R. China\\
              \email{zwliu@ynao.ac.cn}
\and
Key Laboratory for the Structure and Evolution of Celestial Objects, CAS, Kunming 650216, P.R. China
\and
Center for Astronomical Mega-Science, CAS, 20A Datun Road, Chaoyang District, Beijing 100012, P. R. China
\and
E. A. Milne Centre for Astrophysics, Department of Physics \& Mathematics, University of Hull, HU6 7RX, UK}

  \abstract{

The nature of the progenitors of type Ia supernovae (SNe Ia) remains a mystery. Binary systems consisting of a white dwarf (WD) and a main-sequence (MS) donor are potential progenitors of SNe Ia, in which a thermonuclear explosion of the WD may occur when its mass reaches the Chandrasekhar limit during accretion of material from a companion star. In the present work, we address theoretical rates and delay times of a specific MS donor channel to SNe Ia, in which a helium~(He) star~+~MS binary produced from a common envelope event subsequently forms a WD~+~MS system without the He star undergoing mass transfer by Roche lobe overflow. By combining the results of self-consistent binary evolution calculations with population synthesis models, we find that the contribution of SNe Ia in this channel is around $2.0\times10^{-4}\,\mathrm{yr^{-1}}$. In addition, we find that delay times of SNe Ia in this channel cover a range of about $1.0$--$2.6\,\mathrm{Gyr}$, and almost all SNe Ia produced in this way (about $97\,\%$) have a delay time of $\gtrsim1\,\mathrm{Gyr}$. While the rate of SN Ia in this work is about $10\%$ of the overall SN Ia rate, the channel represents a possible contribution to the old population ($1$--$3\,\mathrm{Gyr}$) of observed SNe Ia.

   }

   \maketitle
%

\section{Introduction}
\label{sec:introduction}

Type Ia supernovae (SNe~Ia) are believed to arise from thermonuclear explosions of white dwarfs (WDs) in binary systems \citep{Hoyl1960}. However, the nature of the companion stars and the exact explosion mechanism remain a mystery \citep[e.g. see][for a review]{Hillebrandt2000, Hillebrandt2013, Maoz2014}. Over the past decades, the single-degenerate (SD) and double-degenerate (DD) scenarios have been widely studied as two potential ways to produce SNe Ia. In the SD scenario, the WD accretes matter from a non-degenerate companion star that could be a main-sequence (MS), a sub-giant, a red giant (RG), or a helium (He) star \citep[e.g.][]{Whelan1973, Han2004}, while the DD scenario involves the merger of two WDs \citep[e.g.][]{Iben1984, Webbink1984}. In both cases, material is accreted from the companion star to trigger a thermonuclear explosion while the WD mass approaches the Chandrasekhar limit \citep[but see also][]{Pakmor2010, Pakmor2011, Pakmor2012}. Moreover, it is also possible to trigger an SN Ia via the so-called double detonation scenario when the WD is below the Chandrasekhar mass. The  WD accretes He material from an He-rich companion (either an He star or an He WD) to trigger a detonation in the accreted He shell when this layer reaches a critical value, and this detonation further triggers a second core detonation by compressional heating to cause the SN Ia explosion (e.g. \citealt{Taam1980, Woosley1986, Bildsten2007, Fink2010, Shen2010, Sim2010, Kromer2010, Gronow2020}). Recently, some other progenitor scenarios have also been proposed for SNe Ia, such as the core-degenerate model \citep[e.g.][]{Ilkov2012, Ilkov2013, Kashi2011}, the WD-WD head-on collision model \citep[e.g.][]{Raskin2009, Rosswog2009, Thompson2011, Kushnir2013}, and the `Spin-up-Spin-down' SD model \citep[e.g.][]{Di-stefano2011, Justham2011, Hachisu2012}.  

In the present study we concentrate on the MS donor channel within the SD scenario, that is to say the WD~+~MS channel (sometimes known as the supersoft channel). In the standard WD~+~MS channel, three evolutionary paths have generally been considered for the formation of WD~+~MS binary systems for SNe Ia (e.g. \citealt{Yungelson2000,Ruiter2009,Meng10,Wang2010,Claeys2014, Liu2015}).

First, in the HG (or RGB)\ channel, the primary star fills its Roche lobe as it evolves to the Hertzsprung gap (HG) or red-giant branch (RGB, sometimes called the first giant branch) phase. The system then undergoes a common envelope (CE) event as a result of unstable mass transfer, producing a He~star~+~MS system. This He star subsequently overfills its Roche lobe and stably transfers mass to its companion MS star. A WD~+~MS system is formed, which eventually causes an SN Ia when the MS star transfers material back to the WD and increases its mass to the Chandrasekhar mass.

Second, in the early asymptotic giant branch (EAGB)\ channel, the binary system first experiences Roche lobe overflow (RLOF) when the primary reaches the EAGB. As the mass transfer is dynamically unstable, a CE event results. Ejection of the CE produces a He~RG~+~MS system rather than a core He-burning star. When the He-RG fills its Roche lobe, stable mass transfer takes place, forming a WD~+~MS system. As with the above channel, mass transfer from the MS companion back to the WD triggers the SN~Ia explosion.

Thirdly, in the thermally pulsing asymptotic giant branch (TPAGB)\ channel, if the binary is sufficiently wide, the primary can reach the TPAGB stage before mass transfer takes place. The resulting phase of unstable mass transfer results in a CE event, the end product of which is a WD~+~MS system.


The HG (or RGB) and EAGB channels have also been called the ``He-enriched MS donor scenario'' by \citet{Liu2017} because a large amount of He-rich material is deposited onto the MS companion star by mass transfer as the He~MS (or He RG~star)~+~MS binary evolves to form a WD~+~MS system. As a consequence, the MS companion star of the WD~+~MS system is significantly He-enriched \citep[see also][]{Liu2018}. We note that previous binary population synthesis studies have found that the HG (or RGB) channel is the most significant route for producing SNe Ia amongst the three evolutionary channels mentioned above \citep[e.g. see][]{Wang2010, Meng10}.

In this work we focus on studying a particular channel for producing WD~+~MS systems. The binary evolutionary path of this channel is depicted in Fig.~\ref{Fig:chart}. Similar to that of the above HG (or RGB)\ channel, a He~star~+~MS system is produced as a result of a CE event. However, instead of experiencing RLOF mass transfer to form a carbon-oxygen (CO)~WD~+~MS system, the primary (i.e. the He star) of these systems directly  evolves into a CO~WD without overfilling its Roche lobe \citep[see also][their Appendix A.1.2]{Toonen2014} Subsequently, this CO~WD accretes hydrogen-rich material from the MS companion, increasing its mass until the Chandrasehkar limit is reached and an SN~Ia explosion is triggered. In the present work, we call this channel 'the non-interacting He stars channel'. While the channel is included in binary population synthesis (BPS) codes along with other channels involving He stars, its role is rarely discussed (and never in detail) when describing the standard MS donor scenario for SNe Ia. The main goal of this work is to address the theoretical rates and delay times of SNe Ia in this channel by performing self-consist binary evolution and population synthesis calculations.

\section{Binary evolution calculation}
\label{sec:binary}

\begin{figure}
   \centering
   \includegraphics[width=0.48\textwidth, angle=0]{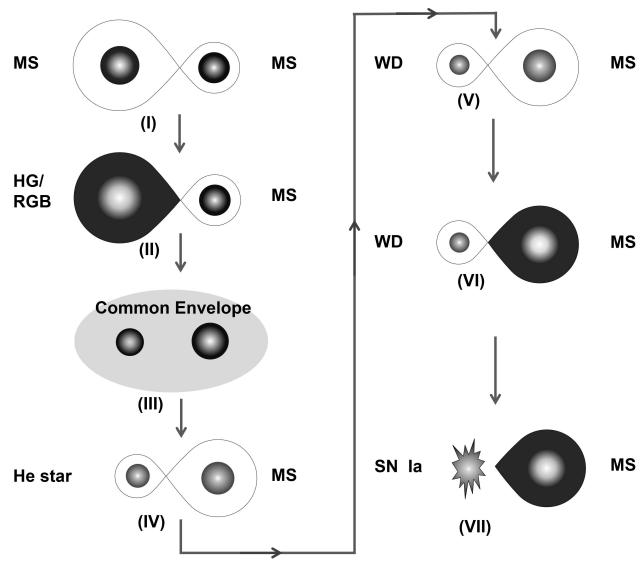}
   \caption{Binary evolutionary path of the studied MS donor channel for SNe Ia in the present work. The He star of the binary system at stage IV does not overfill its Roche lobe to undergo the RLOF mass transfer before it evolves to be a CO~WD at stage V.}
\label{Fig:chart}%
\end{figure}

\begin{figure}
   \centering
   \includegraphics[width=0.49\textwidth, angle=360]{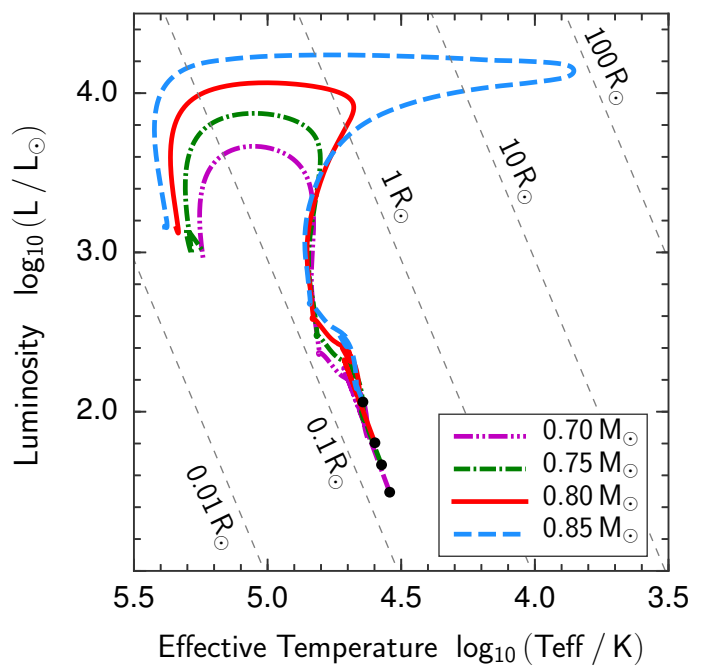}
   \caption{Hertzsprung-Russell (H-R) diagram showing the evolutionary tracks of the primary stars (i.e. He stars) of He~star~+~MS systems in our detailed stellar evolution calculations using the {\sc STARS} code. Different curves show the evolution of the He star with masses of $0.70\,M_{\sun}$, $0.75\,M_{\sun}$, $0.80\,M_{\sun}$, and $0.85\,M_{\sun}$. The starting point of our calculations is given by black dots. The grey dashed lines represent constant radius lines on the H-R diagram.}
\label{Fig:hr}%
\end{figure}

\begin{figure}
  \begin{center}
    {\includegraphics[width=0.49\textwidth, angle=360]{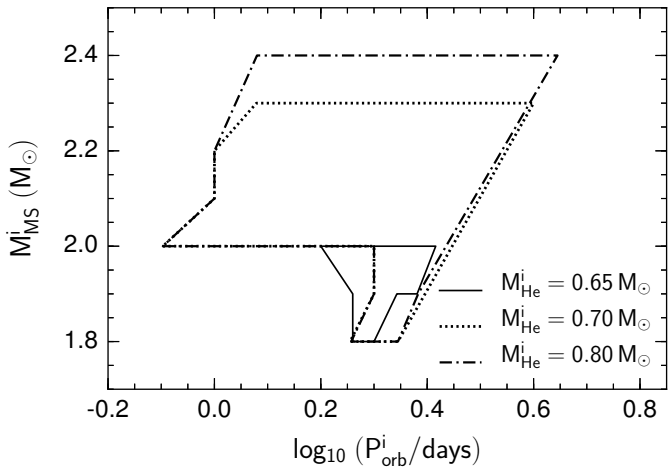}}
    \caption{Regions in the orbital period-secondary mass plane for He~star~+~MS binary systems (stage IV of Fig.~\ref{Fig:chart}) that successfully produce SNe Ia based on our consistent binary evolution calculations. Different curves correspond to regions for different He star masses.}
\label{Fig:contour}
  \end{center}
\end{figure}

\begin{figure*}
  \begin{center}
    {\includegraphics[width=0.46\textwidth, angle=360]{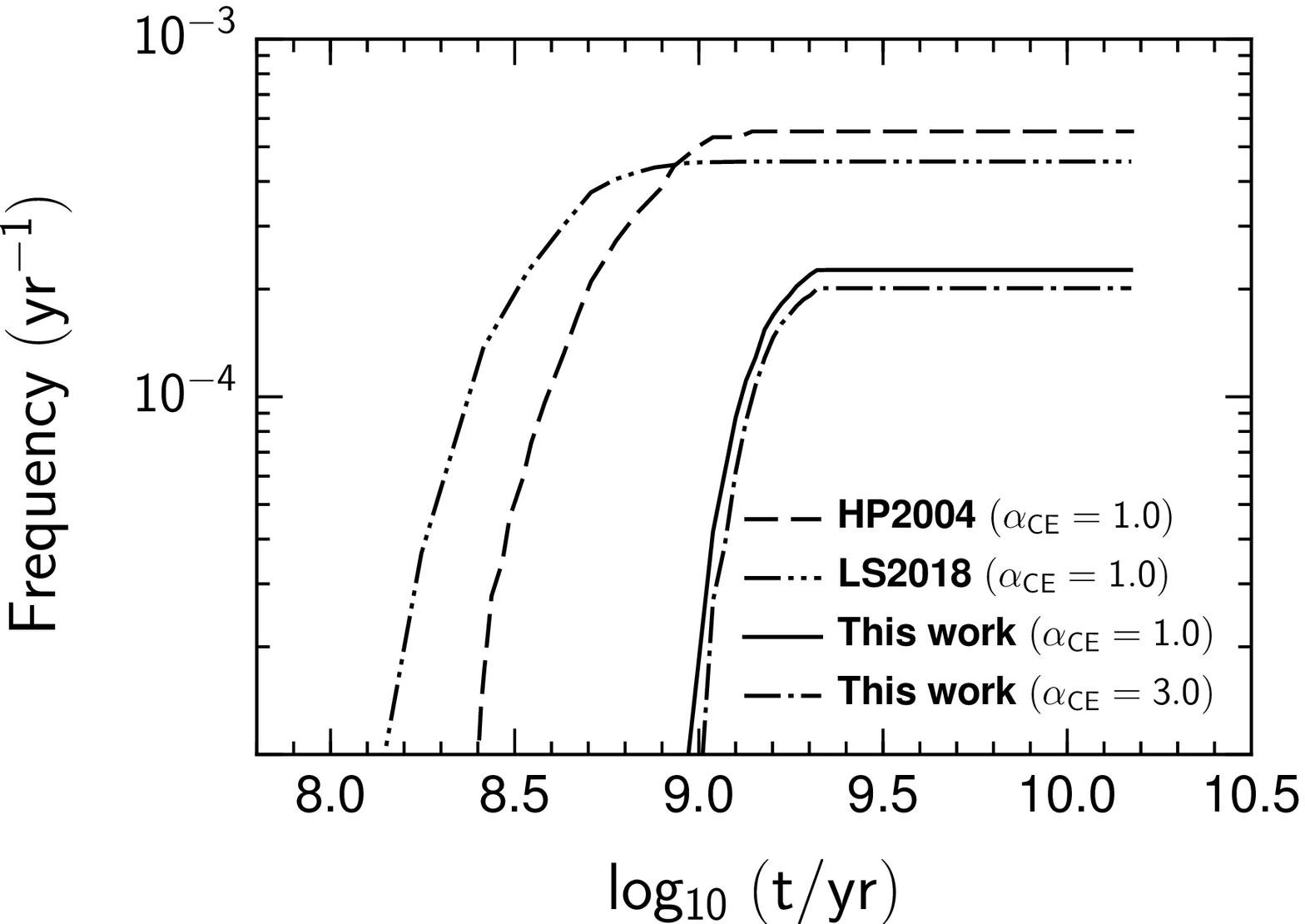}}
    \hspace{8mm}
    {\includegraphics[width=0.46\textwidth, angle=360]{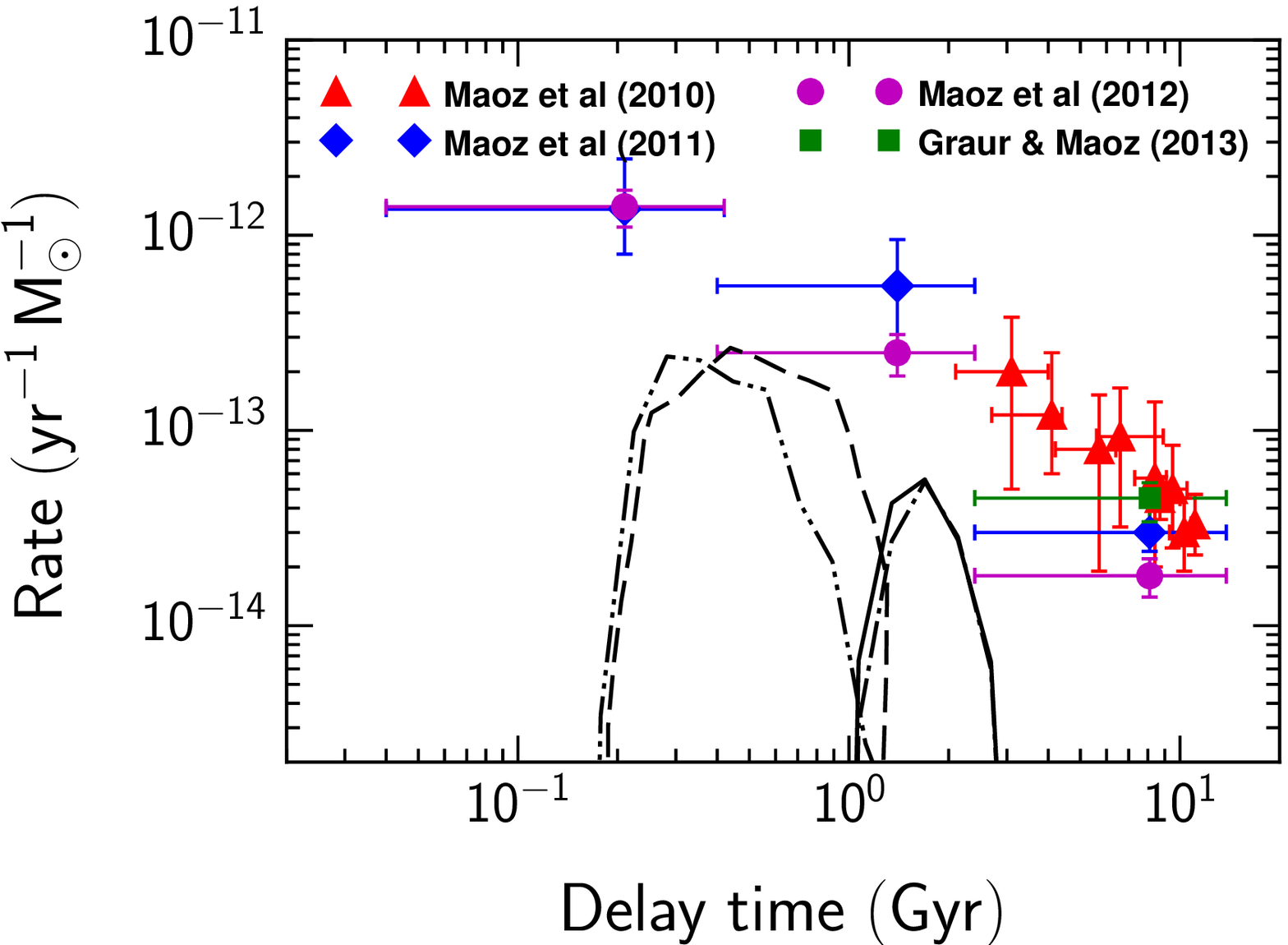}}
    \caption{Left-hand panel: theoretical Galactic rates of SNe Ia as a function of delay time in our studied channel for a constant SFR of $5\,M_{\sun}\,\mathrm{yr^{-1}}$ with the CE ejection efficiency of $\alpha_{\rm CE}=1.0$ (dash-dotted line) and $\alpha_{\rm CE}=3.0$ (solid line). For a comparison, the results of \citet[][i.e. HP2004, see the dashed line]{Han2004} and \citet[][i.e. LS2018, see the double-dotted line]{Liu2018} are also shown in the figure. Right-hand panel: same as the left-hand panel but for the single starburst case of $10^{11}\,M_{\sun}$. Our measured DTD is compared to the observations \citep{Maoz2010, Maoz2011, Maoz2012, Graur2013}.}
\label{Fig:rate}
  \end{center}
\end{figure*}

The method used by \citet[][]{Han2004}, in which the results of self-consistent binary evolution calculations are combined with population synthesis models, is taken to make predictions on theoretical rates and delay times of SNe Ia in the studied channel of the present study (see also \citealt{Liu2018}). Detailed binary evolution calculations are made using the Cambridge stellar evolution code {\sc STARS} \citep{Eggleton1971, Eggleton1972, Pols1995, Stancliffe2009}. Binary evolution for a set of He~star~+~MS systems are self-consistently traced from stage IV to VII (in Fig.~\ref{Fig:chart}) to obtain initial parameter contours at stage~$\mathrm{IV}$, which eventually lead to a successful SN Ia (i.e. $M_{\mathrm{He}}^{\,\mathrm{i}}$, $M_{\mathrm{MS}}^{\,\mathrm{i}}$, and $P_{\mathrm{orb}}^{\,\mathrm{i}}$). Once the He star at stage IV evolves to become a WD and its MS companion starts to overfill its Roche lobe at stage VI, we treat the WD as a point mass rather than solving the detailed structure of the WD. We calculated the mass growth rate of the WD (i.e. $\dot{M}_{\mathrm{WD}}$) from stage VI to VII based on the models of \citet[][for hydrogen accretion]{Hachisu1999} and \citet[][for He accretion, see also \citealt{Kato2004}]{Kato1999}. We refer to \citet{Liu2018} for a detailed description (their Sect.~2.1). We assumed that the accreting WD triggers an SN Ia explosion (at stage VII) when its mass is close to the Chandrasekhar-mass-limit, which we took as $1.4\,M_{\sun}$ in our detailed binary evolution calculations.

We performed self-consistent binary evolution calculations for around 15600 binary systems consisting of a He star and an MS companion star by covering: a range of an initial mass of the primary (He) star of $M_{\mathrm{He}}^{\mathrm{\;i}}=0.3$--$0.9\,M_{\sun}$ in steps of 0.05 $M_\sun$; a range of the initial mass of the secondary (MS) star of $M_{\mathrm{MS}}^{\mathrm{\;i}}=1.6$--$4.0\,M_{\sun}$ in steps of $0.05\,M_\sun$; and a range of initial orbital period of $P_{\mathrm{orb}}^{\mathrm{\;i}}=0.6$--$10\,\mathrm{days}$ in steps of 0.2 days. 

Figure~\ref{Fig:hr} shows the evolution of He stars with an initial mass between  $0.70\,\Msun$ and $0.85\,\Msun$. The 0.85$M_\sun$ model evolves to become a He giant, with a radius of nearly 100$R_\sun$. Such systems always overfill their Roche lobes for the period range we consider. These systems would transfer material to their companions, and they belong to the He-enriched donor channel \citep{Liu2018}; we do not consider them further. However, for He stars of 0.80$M_\sun$ and below, such expansion does not occur. These stars may remain within their Roche lobes and evolve to form CO WDs.

Figure~\ref{Fig:contour} shows the initial parameter space of He~star~+~MS binary systems at stage IV of Fig.~\ref{Fig:chart} in the orbital period-secondary mass plane (i.e. $\mathrm{log_{10}}\,P_{\mathrm{orb}}^{\mathrm{\;i}}$--$M_{\mathrm{MS}}^{\mathrm{\;i}}$) that are capable of producing SN~Ia explosions in our studied MS donor channel. We find that SNe Ia can be produced for He star initial masses in the range  $0.65\,M_{\sun}$ to $0.80\,M_{\sun}$ at stage IV. The boundary for short-period systems in Fig.~\ref{Fig:contour} is constrained by requiring that the systems do not fill their Roche lobes at the beginning of the simulation. Systems with high secondary mass and/or wide initial periods experience dynamically unstable mass transfer when the secondary overfills its Roche lobe and eventually lead to CE events. For systems with low initial secondary mass, the donor star is not massive enough and/or the mass-retention efficiency is not sufficiently high for the WD to be able to reach the Chandrasekhar limit.

\section{Population synthesis calculation}
\label{sec:bps}

Population synthesis calculations are performed by using a rapid binary evolution code \citep{Hurley2000, Hurley2002}. The evolution of $10^{\mathrm{7}}$ binary systems, which are generated in a Monte Carlo way, are followed from their zero-age MS (ZAMS, i.e. stage I of Fig.~\ref{Fig:chart}) phase to the formation of He~star~+~MS systems (stage IV of Fig.~\ref{Fig:chart}). Once a binary system in the rapid binary calculations evolves to stage~$\mathrm{IV}$ and falls into the initial parameter contours given by the above self-consistent binary evolution calculations (shown in Fig.~\ref{Fig:contour}), we assume  a successful SN Ia event is eventually produced. For initial inputs and important assumptions of our population synthesis calculations, we refer to \citet{Liu2018} for a detailed description. Here, we only briefly  summarize the five assumptions as follows.

 First, when a CE event occurs, we ejected the entire envelope using the standard $\alpha$-prescription \citep{Paczynski1976, Webbink1984} as follows:

\begin{equation}
      \alpha_{\mathrm{CE}}\left(\frac{GM_{\mathrm{core}}M_{\mathrm{acc}}}{2a_{\mathrm{f}}}-\frac{GMM_{\mathrm{acc}}}{2a_{\mathrm{i}}}\right)=\frac{GMM_{\mathrm{env}}}{\lambda R}
, \end{equation}

where: $M$, $M_{\mathrm{core}}$, $M_{\mathrm{env}}$, and $M_{\mathrm{acc}}$ represent the total mass, core mass, and envelope mass of the donor star, and the mass of the accretor, respectively; $R$ is the radius of the donor star; $a$ is the separation of the binary, and the subscripts `i' and `f' refer to the initial and final values; $\lambda$ is a structure parameter of the envelope that depends on the evolutionary stage of the donor, which is set to be $\lambda=0.5$ in our calculations; and $\alpha_{\mathrm{CE}}$ is the CE ejection efficiency. In this work, we set $\alpha_{\mathrm{CE}}=1.0$ and/or $\alpha_{\mathrm{CE}}=3.0$. This means that the combination of $\alpha_{\mathrm{CE}}$ and $\lambda$ (i.e. $\alpha_{\mathrm{CE}}\lambda=0.5$, 1.5) is set to be a free parameter to investigate its influence on the results\footnote{This will be discussed further in Sect.~\ref{sec:uncertainties}.}.

Second, a circular orbit is assumed for all binary systems. Third, the primordial primary is generated according to the formula in \citet{Eggleton1989}, which adopted a simple approximation to the initial mass function (IMF) in \citet{Miller1979}.

Fourth, a uniform mass-ratio ($q$=$M_{\mathrm{2}}/M_{\mathrm{1}}$) distribution is adopted  \citep{Mazeh1992} and is uniformly distributed in the range~$[0,1]$ (i.e. $n(q)=1$ for $0<q\leq1$).
Fifth, for the separation $a$ of the binary systems, a constant distribution in $\log\,a$ is assumed for wide binaries, while $a$ falls off smoothly for close binaries \citep{Han1995}:
\begin{equation}
a\cdot n(a)=\left\{
\begin{array}{lcl}
\alpha_{\rm sep}(a/a_{\rm 0})^{1.2} & & a\leq a_{\rm 0} \\
\alpha_{\rm sep} & & a_{\rm 0}<a<a_{\rm 1} \\
\end{array}\right.
, \end{equation}
where $\alpha_{\rm sep}$=$0.07$, $a_{\rm 0}$=$10\,R_{\odot}$, and $a_{\rm 1}$=$5.75\times 10^{\rm 6}\,R_{\odot}$=$0.13\,{\rm pc}$. This setup implies an equal number of wide binary systems per logarithmic interval and gives approximately 50 percent of binary systems with an orbital period $\lesssim100$\,yr.


It is important to keep in mind that the results of population synthesis calculations depend on uncertain initial conditions and assumptions. Examples include (but are not limited to): the current the star-formation rate (SFR), the IMF, and the CE evolution. These are all weakly constrained by current studies \citep[e.g. see][for a detailed discussion]{Claeys2014}. 


\section{Rates and delay times of SNe Ia}
\label{sec:results}

Using the above method, we computed the theoretical birth rates and delay times for SNe~Ia formed by our channel. Our predictions of the evolution of SNe Ia rates  are made by assuming either a constant SFR of $5.0\,M_{\sun}\,\rm{yr^{-1}}$ (left-hand panel) or a single starburst of $10^{11}\,M_{\sun}$ (right-hand panel\footnote{In reality, however, the star-formation history of the Milky Way may be different from a constant SFR of $5.0\,M_{\sun}\,\rm{yr^{-1}}$ or a single starburst of $10^{11}\,M_{\sun}$. For example, \citet{Kubryk2015} suggest that a more realistic SFR of the Milky Way is composed of different bulge and disc components \citep[see also][]{Maoz2017}.}). The results are given in Fig.~\ref{Fig:rate}. By adopting different CE ejection efficiencies (i.e. $\alpha_{\rm CE}=1.0$ and/or $\alpha_{\rm CE}=3.0)$, it is found that the Galactic rate in this channel is around $2.0$--$2.2\times10^{-4}\,\rm{yr^{-1}}$. Compared with the observationally inferred rate of $2.84\pm0.60\times10^{-3}\,\rm{yr^{-1}}$ \citep[e.g.][]{van-den-Bergh1991, Cappellaro1997, Li2011a, Li2011b, Maoz2012PASA}, our prediction is lower by about one order of magnitude. A comparison between the results of this work and those of the standard WD~+~MS channel obtained by previous studies by \citet[][i.e. the dashed curve]{Han2004} and \citet[][i.e. the double-dotted curve]{Liu2018} is also presented in Fig.~\ref{Fig:rate}. It is shown that the Galactic SN Ia rates from the standard WD~+~MS channel are about $0.6$--$1.2\times10^{-3}\,\rm{yr^{-1}}$, which is higher than our results by a factor of $3$--$6$. In addition, the left-hand panel of Fig.~\ref{Fig:rate} shows that the SN Ia rate keeps flat after a sharp rise. This is because the Galactic SN Ia rate is the convolution of the delay-time distribution (DTD) with the star-formation history \citep{Greggio2008, Wang2010}, and it is only related to the DTD when the SFR is constant, leading to the SN Ia rate being constant after the maximum delay times.

Figure~\ref{Fig:rate} presents the DTD of the studied channel for the single starburst case of $10^{11}\,\Msun$ (see right-hand panel). We find that the delay times of SNe Ia in our channel cover a range of around $1.0$--$2.6\,\mathrm{Gyr}$ after the burst (right-hand panel of Fig.~\ref{Fig:rate}). Observations show that SNe Ia consist of a young population (<$100\,\mathrm{Myr}$) and an old population (>$1\,\mathrm{Gyr}$). It has been suggested by previous studies that the young SNe Ia are generated from accreting WDs with a He-star companion and the old population comes from the WD~+~RG channel. The WD~+~He~star channel gives SN Ia rates of $3\times10^{-4}\,\mathrm{yr^{-1}}$ and contributes SNe Ia with delay times of $\lesssim100\,\mathrm{Myr}$, while the WD~+~RG channel predicts Galactic birth rates of $3\times10^{-5}\,\mathrm{yr^{-1}}$ and delay times of $\gtrsim3\,\mathrm{Gyr}$ \citep[e.g. see][]{Ruiter2009, Wang2010, Claeys2014}.

Our predicted DTD is also compared with the observed DTD obtained by different studies \citep[e.g. see][]{Maoz2010, Maoz2011, Maoz2012, Maoz2017, Graur2013}. The observed DTD covers a wide range of delay times from a few $\times\,10\,\mathrm{Myr}$ to about $10\,\mathrm{Gyr}$. Again, a comparison between our results (solid curve for $\alpha_{\rm CE}=1.0$ and dash-dotted curve for $\alpha_{\rm CE}=3.0$) and those from the standard MS donor channel of \citet[][dashed curve]{Han2004} and the He-enriched donor channel of \citet[][double-dotted curve]{Liu2018} is also shown in right-hand panel of Fig.~\ref{Fig:rate}. It shows that the standard MS donor channel is expected to contribute to SNe Ia with intermediate delay times of about $100\,\mathrm{Myr}$--$1\,\mathrm{Gyr}$.
However, in the present work, SNe Ia from our non-interacting He star channel cover a range of delay times of about $1\,\mathrm{Gyr}$--$2.6\,\mathrm{Gyr}$, and almost all SNe Ia (about $97\,\%$) have a delay time of $\gtrsim1\,\mathrm{Gyr}$.  Thus, our channel mainly contributes to the old population  ($1$--$2.6\,\mathrm{Gyr}$) of SNe Ia. On the other hand, it is important to point out that our results, $1$--$2.6\,\mathrm{Gyr}$, bridge the DTD between the standard MS donor channel, $100\,\mathrm{Myr}$--$1\,\mathrm{Gyr}$, and the WD~+~RG channel, $\gtrsim3\,\mathrm{Gyr}$, although they can only cover a narrow range of the observed DTD.

By considering a circumbinary disc that extracts the orbital angular momentum from the binary through tidal torques in their calculations for the standard MS donor channel, \citet{Chen2007} suggested that the standard MS donor channel could also contribute to the old SNe Ia of $1$--$3\,\mathrm{Gyr}$. However, our understanding of the circumbinary disc is still quite poor. In addition, from current studies on the SD scenario, it seems to be difficult to cover the whole range of delay times of observed SNe Ia with a single non-degenerate donor channel. This may indicate that a combination of different non-degenerate donor channels would probably be needed if SNe Ia are indeed produced from the SD scenario. It is also interesting to note that the expected DTD from the DD scenario by population synthesis studies seems to provide a better match with the whole range of the observations \citep[e.g. see][]{Ruiter2009, Ruiter2011, Claeys2014, Yungelson2017}, although it is still too early to conclude that SNe Ia are likely to mainly be produced from the DD scenario \citep[see][for a recent review]{Maoz2012, Livio2018}.

\section{Uncertainties of the results}
\label{sec:uncertainties}

In order to compute the evolution of a large number of systems, population synthesis studies have to simplify some complex physics. Even with detailed binary calculations, simplifications have to be made in order for the computational time need to remain tractable. The choices made  must necessarily affect the results, and so we now discuss some of the possible uncertainties.

In this work, the methods of \citet{Hachisu1999} and \citet{Kato1999} are used to describe the retention efficiency of hydrogen and He onto the WD, respectively. This is similar to methods used by \citet{Han2004}. However, the exact mass-retention efficiency of accreted material onto a WD is poorly constrained by current studies because the novae phase \citep{Yaron2005} and the phase with the high-mass transfer rates \citep{Hachisu1999} are still fairly uncertain. Different mass-retention efficiencies have been used in various BPS studies, and it has been suggested that they could strongly affect the outcomes of these calculations, such as SN Ia rates and the DTD. For instance, \citet{Bours2013} find that predicted SN Ia rates could vary by a factor of up to more than 100 if different retention efficiencies are adopted  \citep[see also][]{Ruiter2014, Piersanti2014, Toonen2014}. In particular, we note that no SD binaries were found to evolve into an SN Ia in the calculations of \citet{Bours2013} when the retention efficiency of \citet{Yungelson2010} was used. A detailed study of the effect of different retention efficiencies on the BPS results can be found in \citet{Bours2013} and \citet{Ruiter2014}. Moreover, it has been found that the initial temperature of the WD can affect the retention efficiency \citep[e.g.][]{Yaron2005, Chen2019}, although our calculations do not take this effect into account since the accreting WD is treated as a point mass. Future studies are encouraged to provide a more realistic prescription of retention efficiency based on a rigorous treatment of radiative transfer and with the inclusion of the effect of the WD's initial temperature.

The details of how CE evolution works are perhaps some of the most uncertain in all of binary star physics \citep{Ivanova2013}. We have used the $\alpha$-prescription \citep{Paczynski1976, Webbink1984} to calculate the outcome of the CE phase in this work. The combination of $\alpha_{\mathrm{CE}}$ and $\lambda$ (i.e. $\alpha_{\mathrm{CE}}\lambda$) is set to be 0.5 and 1.5 to investigate its influence on the results (see also \citealt{Yungelson2017}). The $\gamma$-algorithm has also been used for the CE phase \citep{Nelemans2000}. It has been shown that the SN Ia rates could be changed by a factor of two if the $\gamma$-formalism is used to replace the $\alpha$-prescription \citep{Bours2013}, and the different values of $\alpha_{\mathrm{CE}}\lambda$ can lead to a variation in the SD SN Ia rates by up to an order of magnitude \citep{Claeys2014}. In addition, we used a fixed value for $\lambda$, which expresses how tightly bound the star's envelope is. This quantity is not constant for all stars and is expected to vary with stellar mass, envelope mass, luminosity, and evolutionary stage; \citet[][see their Appendix A]{Claeys2014} therefore also used a fitting formula to calculate $\lambda$ \citep[see also][]{Dewi2000, Izzard2004}. We refer  to \citet{Claeys2014} for a more detailed discussion of different treatments of the CE phase and their effects on the BPS results.

\section{Conclusion and summary}
\label{sec:results}

We have investigated the formation of SN~Ia via an MS donor channel where the He star does not interact with its MS companion prior to it forming a CO WD.  This is different from that of the HG (or RGB) channel of the standard MS donor scenario, in which RLOF mass transfer is expected to happen during the formation of WD~+~MS systems from the evolution of He~star~+~MS systems (see Sect.~\ref{sec:introduction}). We computed theoretical rates and delay times of this channel by combining the results of self-consistent binary evolution calculations with population synthesis models.  The results of this work are summarized as follows.

   \begin{enumerate}
      \item[1)] We find He~star~+~MS binary systems at stage~IV of Fig.~\ref{Fig:chart}, which can eventually lead to an SN Ia having a He star mass of $M_{\mathrm{He}}^{\mathrm{\;i}}=0.65$--$0.80\,M_{\sun}$, an MS donor mass of $M_{\mathrm{MS}}^{\mathrm{\;i}}=1.8$--$2.4\,M_{\sun}$, and an initial orbital period of $P_{\mathrm{orb}}^{\mathrm{\;i}}=0.6$--$5.0\,\mathrm{days}$ based on our self-consistent binary evolution calculations (see Fig.~\ref{Fig:contour}). 
      
      \item[2)] By assuming a constant SFR of $5\,M_{\sun}\,\rm{yr^{-1}}$, we find that the Galactic SN Ia rate in the specific MS donor channel of this work is about $2\times10^{-4}\,\rm{yr^{-1}}$, which is lower than the observationally inferred rate of $2.84\pm0.60\times10^{-3}\,\rm{yr^{-1}}$ \citep[e.g.][]{Li2011a, Li2011b, Maoz2012PASA} by about one order of magnitude. In addition, our results are lower than the predicted SN Ia rates ($0.6$--$1.2\times10^{-3}\,\rm{yr^{-1}}$, e.g. see \citealt{Han2004}) of the standard MS donor channel by a factor of $3$--$6$. 
      
      \item[3)] The SN Ia rates from this work are lower that of the standard MS donor channel \citep{Han2004, Liu2018}, but they are higher than the prediction of the WD~+~RG channel of $3\times10^{-5}\,\rm{yr^{-1}}$ by one order of magnitude. Therefore, this channel cannot be ignored in theoretical studies on rates and delay times of SD SNe Ia.

      \item[4)] The DTD of SNe Ia from our studied MS donor channel covers a range of $\sim1.0$--$2.6\,\mathrm{Gyr}$ after the burst as shown in right-hand panel of Fig.~\ref{Fig:rate}, and almost all SNe Ia (about $97\%$) from this work have a delay time of $>1\,\mathrm{Gyr}$. It is therefore concluded that the studied MS donor channel of this work suggests a new possible way to contribute to the old population of SNe Ia ($\gtrsim1\,\mathrm{Gyr}$). This is different from the DTD of the standard MS donor channel, which is expected to contribute to SNe Ia with delay times of $100\,\mathrm{Myr}$--$1\,\mathrm{Gyr}$ \citep[e.g.][]{Han2004, Ruiter2009, Liu2018}. 
      
       \item[5)] Our results ($1.0$--$2.6\,\mathrm{Gyr}$) provide a potential way to bridge the DTD between the standard MS donor channel ($100\,\mathrm{Myr}$--$1\,\mathrm{Gyr}$, \citealt{Han2004})  and the WD~+~RG channel ($\gtrsim3\,\mathrm{Gyr}$, \citealt{Wang2010}).

\end{enumerate}

\begin{acknowledgements}
We would like to thank the anonymous referees for their reviews. We thank Robert G. Izzard and Hai-Liang Chen for their fruitful discussions. ZWL is supported by the National Natural Science Foundation of China (NSFC, No. 11873016) and the Chinese Academy of Sciences. RJS is supported by STFC, through the University of Hull Consolidated grant ST/R000840/1, and acknowledges use of the University of Hull HPC Facility, Viper.
\end{acknowledgements}

\bibliographystyle{aa}

\bibliography{ref}

\end{document}